\documentclass[aps,prl, twocolumn, groupedaddress]{revtex4-2}
\usepackage{lipsum}
\usepackage{mathtools}
\usepackage{graphicx}
\usepackage{dcolumn}
\usepackage{amsmath}    
\usepackage{amssymb}
\usepackage{bm}
\usepackage{hyperref}
\usepackage{latexsym}
\usepackage{verbatim}
\usepackage[normalem]{ulem}
\usepackage{color}
\usepackage[caption=false]{subfig}
\def\beq{\begin{equation}}
\def\eeq{\end{equation}}

\def\beq{\begin{equation}}                          
\def\eeq{\end{equation}}                          
\def\bea{\begin{eqnarray}}                          
\def\eea{\end{eqnarray}}

\DeclareRobustCommand{\uvec}[1]{{%
  \ifcsname uvec#1\endcsname
     \csname uvec#1\endcsname
   \else
    \bm{\hat{\mathbf{#1}}}%
   \fi
}}
\preprint{}
\usepackage[normalem]{ulem}

\preprint{}
\begin{document}
\title{Ion Permeation in Nanoscale Films: Fundamental Limitation and Evaluation of Dielectric Properties}

\author{Jay Prakash Singh$^{1}$}
\email{jay.singh@campus.technion.ac.il}
\author{Konstantin I. Morozov $^{1}$}
\email{mrk@technion.ac.il}
\author{Viatcheslav Freger$^{1,2,3}$}
\email{vfreger@technion.ac.il}

\affiliation{$^{1}$ Wolfson Department of Chemical Engineering, Technion Israel Institute of Technology, Haifa, Israel.}
\affiliation{$^{2}$ Grand Technion Energy Program, Technion Israel Institute of Technology, Haifa, Israel.}
\affiliation{$^{3}$ Grand Water Research Institute, Technion Israel Institute of Technology, Haifa, Israel.}

\date{\today}
\begin{abstract}
Nanoscale films play a central role in biology and osmotic separations. Their water/salt selectivity is often regarded as intrinsic property, favoring thinner membranes for faster permeation. Here we highlight and quantify a fundamental limitation arising from the dependence of ion self-energy on film thickness, governed by its ratio to Bjerrum length. The resulting relation factors out this dependence from intrinsic ion permeability, which agrees well with available data and enables evaluation of dielectric properties of ultrathin films, advancing understanding of ion transport in membranes.

\end{abstract}
\maketitle

\section{Introduction}
Thin films play a central role across both biological systems and modern technologies, particularly in water desalination and selective ion transport \cite{fane_wang2015synthetic, park_elimelech2017maximizing, freger_ramon2021polyamide}. In biological membranes, ultrathin lipid bilayers efficiently block non-selective transport through specialized pore proteins \cite{fujiyoshi_Agre2002structure, maffeo_aksimentiev2012modeling, song_manishkumar2020artificial}. Today's synthetic membranes rival this performance, with nanofabrication enabling precise control at the nanometer scale \cite{gu_stafford2013molecular, yang_darling2018atomic, bruening2008creation, regenspurg_deWos2024polyelectrolyte}. Such films promise revolutionary gains in permeance and energy efficiency for next-generation separations \cite{nunes_ramon2020thinking, epsztein2020towards}.

The drive toward ever-thinner membranes raises a profound fundamental question: does there exist a fundamental limit where further reductions in thickness no more improves, or even degrades, membrane performance? For small neutral permeants like water, permeability is governed by local solute-membrane interactions (partitioning) and mobility (diffusivity), making it an intrinsic, weakly thickness-dependent material property down to molecular dimensions \cite{mulder2012basic, paula1996permeation}. However, charged solutes experience long-range electrostatic interactions with their image charges in the surrounding high-dielectric aqueous phases. Consequently, the energy barrier a permeating ion must overcome depends not only on the local chemistry and nanostructure but also on the macroscopic geometry and, specifically, the membrane thickness. This introduces a fundamental trade-off: while thinning a membrane shortens the diffusion path and increases water permeability, it simultaneously alters the energy landscape in a way that modifies and potentially compromises ion-water selectivity. Thus, the classical paradigm of thickness reduction as beneficial for membrane performance may fail for ions.

\begin{figure}[h!]
    \centering
    \includegraphics[width=0.4\textwidth]{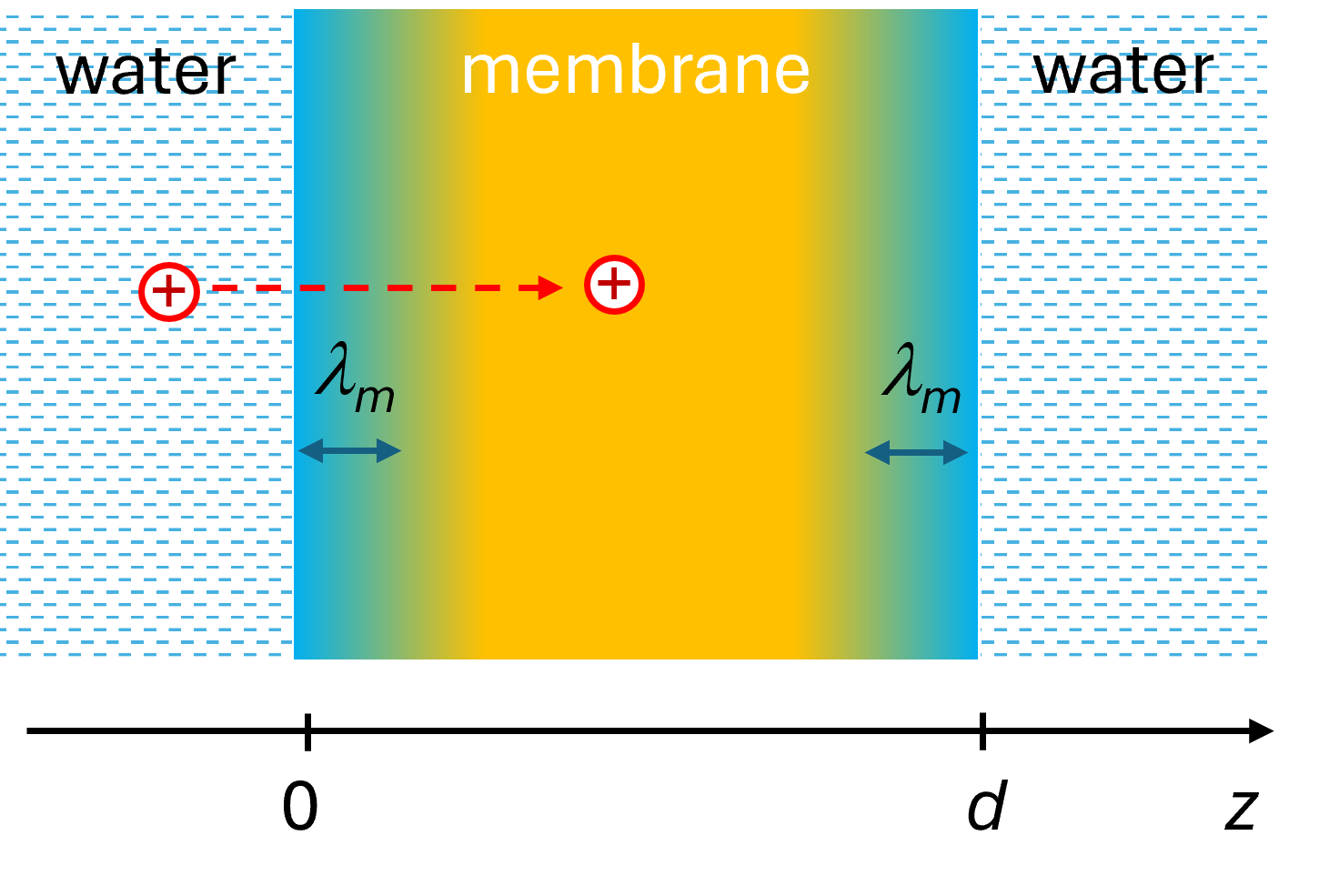} 
    \caption{Sketch of an ion crossing a planar film (membrane) of thickness $d$. $z$ is the thickness coordinates, and $\lambda_m$ represent the characteristic thickness of regions adjacent to the interfaces where ion exclusion weakens substantially.} 
    \label{fig:sketch}         
\end{figure}

In his  seminal paper, Parsegian calculated the electrostatic ``self-energy'' barrier for a point charge crossing a low-dielectric slab of thickness $d$ and dielectric constant $\epsilon_m$ immersed in a high-dielectric medium ($\epsilon_w$)\cite{parsegian1969energy}.  In the middle plane of a film, the self-energy (scaled by thermal energy $k_B T$) is reduced compared with an infinitely thick film by
\begin{equation} \label{Parsegian}
U_m = -\frac{\lambda_m}{d} \ln(1{+} \Delta) \approx -\frac{\lambda_m}{d} \ln 2,
\end{equation}
where $\lambda_m = e^2/(4\pi \epsilon_0 \epsilon_m k_B T)$ is the Bjerrum length inside the membrane, $e$ the electron charge, $\epsilon_0$ the vacuum electrical permittivity, and $\Delta = (\epsilon_w-\epsilon_m)/(\epsilon_w+\epsilon_m)$. The last approximate expression holds for $\epsilon_m \ll \epsilon_w$, as common for lipid and polymeric membranes in water. Equation \eqref{Parsegian} clearly shows that the barrier lowers as $d$ decreases and imposes a fundamental limitation on using the thickness as a cue for improving membrane performance. 

However, eq. \eqref{Parsegian} also suggests a curious opportunity to utilize this dependence for evaluating the dielectric properties of membranes, which proves to be non-trivial for ultrathin films. Standard dielectric characterization techniques such as impedance spectroscopy are inapplicable or highly challenging for films  thinner that $ \sim$100 nm, as they are hampered by interfacial polarization, substrate coupling, leakage currents, and parasitic reactions, obscuring the intrinsic dielectric response of the film itself, especially, for hydrated films \cite{freger2007characterization,nahir2005impedance, venkatesh2005overview}. Here, ion or salt permeability offers an attractive alternative, however, deducing dielectric properties directly from measured permeability would be extremely challenging due to insufficient understanding of underlying physics.  Virtually all models of ion transport in membranes critically depend on dielectric constant but it usually combines in an \textit{a priori} unknown model-dependent manner with other characteristics, often uncertain too \cite{yaroshchuk2001non, freger2020ion, epsztein2020towards}. The latter may include ion size (bare or hydrated), pore or free volume cavity size and shape, and various specific interaction within the film that may also be irregular and heterogeneous, e.g., polyamide network in desalination membranes \cite{ kolev2015molecular}. In contrast to these molecular-scale parameters, defining the intrinsic ion permeability of the film material, thickness dependence represents a better defined \textit{macroscopic} effect that can be factored out of the intrinsic permeability in a robust and model-independent manner. Moreover, unraveling mechanisms behind intrinsic permeability may largely benefit from the measured $\epsilon_m$. 

Parsegian's relation indicates this thickness dependence but, unfortunately, does not supply the desired permeability-thickness relation $P(d)$, which requires appropriate integration of the spatially varying potential $U(z)$ across the entire film \cite{paula1996permeation}.
Since $U(z)$ is strongly position-dependent, the permeability depends on $d$ in a more complex manner than \eqref{Parsegian} suggests. Past attempts to derive $P(d)$ often employed simplifying but inaccurate assumptions (e.g., treating water as a perfect conductor) and failed to yield a compact analytical solution \cite{haydon1972ion, paula1996permeation}. To this end, the present paper derives an explicit and easy-to-use analytical relation from the entire electrostatic self-energy profile $U(z,d,\epsilon_m,\epsilon_w)$. As a practical approach, the local molecular-scale effects are captured within a thickness-independent, intrinsic permeability $P_\infty$ viewed as the infinite-thickness limit, multiplied by a thickness-dependent correction factor to yield $P(d)$. Subsequently, we demonstrate and discuss the use and limitations of these relation for analysis of dielectric properties of biological and synthetic membrane films. 

\section{Model Derivation and Approximation}

\textit{Relation between self-energy profile and ion permeability.}
We consider steady-state one-dimensional ion transport across a membrane of thickness $d$, as sketched in Fig.~\ref{fig:sketch}. The ion is subject to a position-dependent electrostatic self-energy $U(z)$ arising from long-range interactions, which is superimposed on a position and thickness-independent intrinsic contribution $U_0$. The latter accounts for local, short-range effects such as ion solvation, steric exclusion, and molecular friction within the membrane \cite{yaroshchuk2001non, geise2014fundamental, freger2020ion}.

Ion transport within the membrane is assumed to be governed by diffusion and migration and is described by the Nernst-Planck equation \cite{freger2020ion}
\begin{equation}
\textbf{J} = -D \left( \nabla C + C \nabla U \right),
\end{equation}
where $D$ is the ion diffusivity, $C(z)$ is the ion concentration, and $\textbf{J}$ is the ion flux. Under steady state conditions, integration of this equation across the membrane thickness relates the ion flux to the concentration difference between the two membrane interfaces and to the spatially averaged Boltzmann factor of the self-energy profile, $\langle e^{U(z)} \rangle$. For symmetric boundary conditions, where the electrostatic potential at the two interfaces is identical, the membrane permeability can be expressed as
\begin{equation}
P(d)=\frac{P_{\infty}}{\left\langle e^{U(z)} \right\rangle}.
\label{P general}
\end{equation}
Here, $P(d)$ is the thickness-dependent permeability and $P_{\infty}=De^{-U_0}$ incorporates only intrinsic membrane-ion interactions, while all thickness dependence of the permeability is governed by $\langle e^{U(z)} \rangle$ \cite{yaroshchuk2001non, geise2014fundamental, freger2018selectivity, epsztein2020towards, freger2023dielectric, bannon2024application}.
\begin{figure}
    \centering  \includegraphics[width=1.0\linewidth]{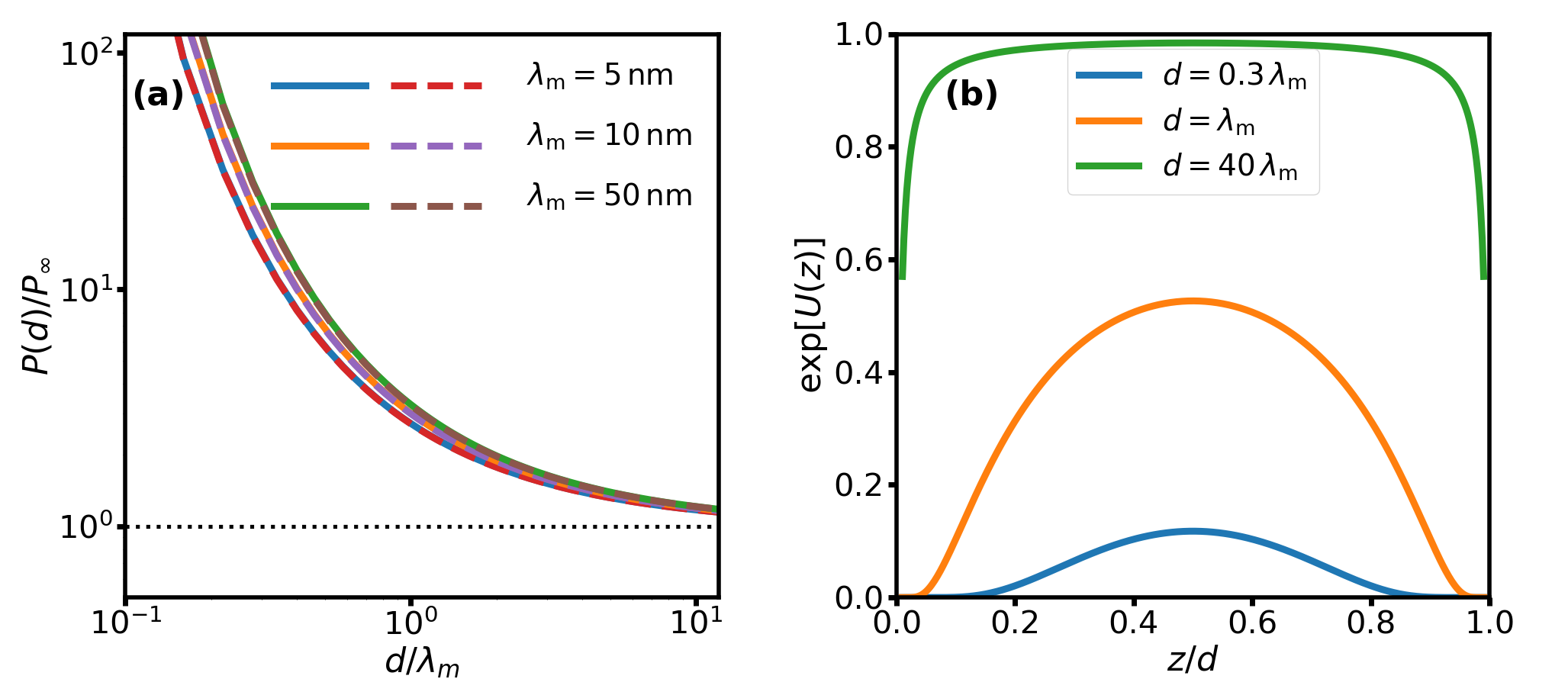}
    \caption{(a) Dependence of ion permeability, scaled with $P_{\infty}$, on the film thickness, scaled with the Bjerrum length $\lambda_m$. The solid and dashed lines represent the full numerical solution and the analytical approximation for different values of $\lambda_m$ (see eqs.~(S10) and (S15) in Supplemental Material).  
(b) Profiles of the thickness-dependent part of ion self-energy $U(z)$ across the film, presented as $e^{U(z)}$, for films much larger ($d = 40\lambda_m$) and equal ($d = \lambda_m$) to the Bjerrum length, and much smaller than the Bjerrum length ($d = 0.1\lambda_m$). The Bjerrum length is 10~nm.
}
 \label{fig:permeability}
\end{figure}

\textit{$U(z)$ and $P(d)$ relations: exact  and approximate expressions.}
To evaluate the thickness dependence encoded in $\langle e^{U(z)} \rangle$, we consider the electrostatic self-energy of a point charge embedded in a planar dielectric slab (membrane), with dielectric permittivity $\epsilon_m$, facing on both sides an aqueous phase with a higher permittivity $\epsilon_w$ (Fig. \ref{fig:sketch}). The profile $U(z)$ is obtained by solving the associated electrostatics problem with appropriate boundary conditions \cite{BatyginToptygin78}. This yields the exact expression for $U(z)$  as a series representation. Through an appropriate approximation, the exact solution may be significantly simplified and converted to a compact analytical expression, as elaborated in the Supplemental Material.  Substituting the approximate self-energy profile into Eq.~\eqref{P general} and subsequent spatial averaging yield the following analytical expression for the thickness dependence of ion permeability,
\begin{equation}
\frac{P(d)}{P_{\infty}} \approx  
(1 {+} \Delta )^{\tfrac{\lambda_m}{d} } \,
\beta\!\left(\frac {\Delta}{2}\frac{\lambda_m}{d} \right),
\label{P approx}
\end{equation}
where $\Delta=(\epsilon_w-\epsilon_m)/(\epsilon_w+\epsilon_m)$ characterizes the dielectric contrast, $\lambda_m$ is the Bjerrum length inside the membrane, and $\beta(x)$ where $\beta(x)=\{ x{e^{x}}[K_1(x)-K_0(x)]\}^{-1}$, and $K_0$ and $K_1$ are the modified Bessel functions of the zero and first order, respectively (see Supplemental Material for detail). Depending on whether $\lambda_m$ or $\epsilon_m$ is fitted, Eq.~\eqref{P approx} can also be recast in terms of the membrane permittivity, using $\lambda_m = \lambda_0/\epsilon_m$, where $\lambda_0 = 56$~nm is the Bjerrum length in vacuum.
As illustrated in Fig. \ref{fig:permeability}a, this result is an excellent approximation, virtually indistinguishable from the exact solution over the relevant range of dielectric contrast and membrane thickness. This plot also highlights the role of the Bjerrum length $\lambda_m$ as the characteristic thickness, below which the permeability increases sharply and selectivity drops. Conversely, above this thickness, the permeability varies weakly and approaches $P_{\infty}$. 

Fig. \ref{fig:permeability}b reveals the basis of such behavior in more detail. The thickness $d = \lambda_m$ signifies the crossover where the self-energy reduction becomes substantial. For $d \gg \lambda_m$, the $U(z)$ profile is close to zero and thus $e^U \approx 1$ nearly everywhere except in thin regions of thickness $\sim\lambda_m$ next to the interfaces. Consequently, $\langle e^U \rangle \approx 1$ and $P \approx P_{\infty}$. Conversely, for $d \ll \lambda_m$, the self-energy is strongly reduced throughout the film, leading to $\langle e^U \rangle \ll 1$ and therefore $P \gg P_{\infty}$. Notably, the flat-top shape of $U(z)$ profiles in our continuum-type model contrasts sharply peaked, $\Lambda$-like
potentials of mean force reported in many molecular dynamics studies fields~\cite{vorobyov2014ion,marsh2001polarity,hopfer1970effect,latorraca2014continuum,chen2021molecular}.
Nevertheless, the present $U(z)$ closely resembles the profiles reported recently by Chen et al. using improved polarizable force-fields, who assigned previously reported $\Lambda$-shaped
potential to the use of non-polarizable force-fields \cite{chen2021molecular}. 

 It is also notable that $U(z)$ diverges near the membrane interfaces, reflecting the strongly reduced electrostatic penalty for placing a charge close to the high-dielectric phase. The corresponding Boltzmann factor $e^{U(z)}$ vanishes at the interfaces thus this singularity is not an issue for calculating $P(d)$. Dissociated surface charges or adsorbed ions may modify $U$ at the boundaries and make it finite \cite{paula1996permeation, haydon1972ion, yaroshchuk2019limiting}. This effect is equivalent to imposing an electric potential bias, which may have a significant effect on the ion distribution in the external aqueous phase yet is usually fairly small (up to a few $k_BT$), compared with the large self-energy within the low dielectric slab. Consequently, it should still weakly affect the spatial average $\langle e^{U(z)} \rangle$ and thus have a minor influence on the resulting dependence $P(d)$. Equation~\eqref{P approx} then  sensibly describes how ion permeability crosses over from a strongly thickness-dependent regime for $d \lesssim \lambda_m$ to a weakly varying regime for $d \gg \lambda_m$ and forms the basis for comparison with experimental data below.

\section{ Discussion and comparison with experiment}
The rapid increase in ion permeability for $d < \lambda_m$ in Fig. \ref{fig:permeability}a would not be observed for uncharged solutes, even as polar as water, since their dipole self-energy shows hence permeability much weaker thickness dependence \cite{paula1996permeation}. This implies that low-dielectric films, such as lipid bilayers or polymer membranes used for desalination, will eventually lose selectivity when made progressively thinner, even if they remain perfectly defect-free. This presents a previously overlooked \textit{fundamental limitation for ion-rejecting membranes}, indicating the approach based on reducing the thickness, even  with defects totally eliminated, still requires a compromise between increasing water permeation rates and maintaining high water/salt selectivity. Eq. \ref{P approx} may help guide such optimization. 

As mentioned above, Eq. \ref{P approx} also offers an intriguing way to deduce dielectric constant of thin films and membranes, a few to a few tens of nanometers thick, based on measured salt or ion permeability. This thickness range is most common in biological cell membranes and is also targeted for separation membranes used in nanofiltration or reverse osmosis \cite{mulder2012basic}. Dielectric characteristics are crucial for modeling ion permeation in such membranes, yet their accurate determination using classical dielectric spectroscopy remains challenging at these length scales \cite{chang2019dielectric}. On the other hand, deducing dielectric properties from $P_{\infty}$ is challenging at present due to insufficient understanding of underlying physics and uncertainties in
ion size, solvation, and steric effects
\cite{yaroshchuk2001non,geise2014fundamental,freger2018selectivity,epsztein2020towards,freger2023dielectric,bannon2024application}.
 In contrast, film thickness can often be measured with high precision, making thickness-dependence of permeability a particularly attractive observable.

\begin{figure}
    \centering
    \includegraphics[width=1.08\linewidth]{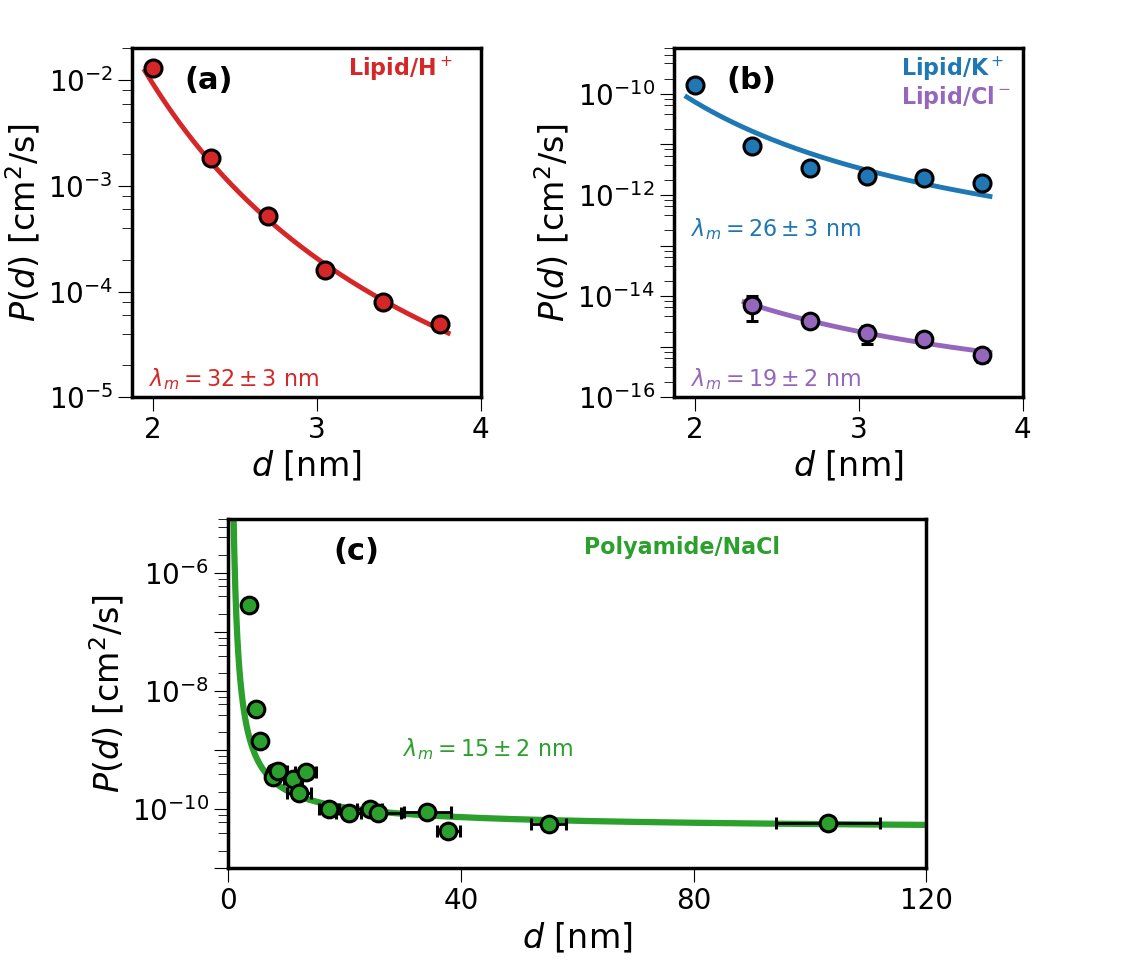}
    \caption{
Thickness dependence of ion permeability.  
(a) Proton permeation through lipid membranes, reproduced from ~\cite{paula1996permeation}.  
(b) Potassium and chloride permeation through lipid membranes from ~\cite{paula1996permeation, paula1998permeation}.  
(c) NaCl permeation through polyamide films, reproduced from~\cite{mulhearn2021thickness}.  
In all panels, symbols show experimental measurements, and solid lines are fits to Eq.~\eqref{P approx}. The reported permeances ($P/d$) were scaled by inverse thickness to obtain the permeability dependence $P(d)$.
}
  \label{fig:proton_lipid}
\end{figure}

We first illustrate this approach for \textit{lipid bilayer membranes}. Fig. \ref{fig:proton_lipid}a, b present permeability data normalized by thickness for protons, potassium and halide ions across lipid bilayer membranes, reported by Paula et al. \cite{paula1996permeation, paula1998permeation}. The thickness corresponds to the hydrophobic core of the bilayer and was systematically varied by changing the hydrocarbon tail length of the lipid molecules. Notably, all experimental points clearly belong to the range   $d<\lambda_m$, which well explains the high sensitivity of the thickness-normalized proton permeability to thickness. The data for proton permeation using Eq. \eqref{P approx} yield a good fit, with $\lambda_m = 32 \pm 3$ nm, reasonably close to $\lambda_m \approx 28$ nm, corresponding to the dielectric constant $\epsilon = 2$ typical of lipid films \cite{haydon1972ion,gramse2013nanoscale}.   

Fig.\ref{fig:proton_lipid}b shows similar fits for potassium and chloride permeation, yielding $\lambda_m = 26 \pm 3$ nm and $\lambda_m = 19 \pm 2$ nm or $\epsilon_m = 2.2 \pm 0.4$ and $\epsilon_m = 2.9 \pm 0.5$, respectively (all errors are 90\% confidence). The slightly higher $\epsilon_m$ values are more consistent with experimental electrostatic force spectroscopy measurements that found $\epsilon_r \sim 3$ for lipid membranes , higher than commonly presumed $\epsilon_r \sim 2$ \cite{gramse2013nanoscale}. However, the fits and uncertainties are clearly not as good as for protons. The discrepancy could be related to 7-11 orders of magnitude smaller permeability, compared with protons, and the resulting larger errors, especially for the least permeable thickest films. We may also speculate that, unlike smaller protons, whose transport within the lipid phase may also be facilitated by water, potassium ions have a more substantial size, between 0.13 nm (bare radius) to about 0.3 nm (hydrated radius), commensurate with the lipid thickness. Thus, potassium ions may (and essentially have to) disrupt the lipid packing to some degree and deviate from the point-charge behavior, as assumed in the present model. Similar reservations may apply to chloride.

Still, with above reservations,
the observed behavior seems to favor a single-ion solution-diffusion mechanism, as argued by
Paula et al.~\cite{gramse2013nanoscale} and, more recently, by Chen et al. \cite{chen2021molecular}, rather than models involving excessive
membrane disruption or deformation by the invading ion \cite{parsegian1969energy, nagle1978lateral, nagle2000structure,gurtovenko2010defect,vorobyov2014ion}. Similarly,
it disfavors cooperative permeation mechanisms that would involve 
simultaneous transport of several ions or ions together with multiple water
molecules disrupting the lipid membrane, such as single-file or wire-like
permeation pathways \cite{malaspina2023few,allen2006ion}, in which cases a weaker thickness dependence is expected.

As another example, we consider in Fig. \ref{fig:proton_lipid}c the NaCl permeability through \textit{synthetic aromatic polyamide films} prepared via molecular layer-by-layer synthesis reported by Mulhearn et al.~\cite{mulhearn2021thickness}. Due to the very wide range of accessible thicknesses, from 6 to 100 nm, the large-$d$ limit $P_\infty$ is clearly identified, and thus could be fixed to keep $\lambda_m$ (or $\epsilon_m$) as the only fitting parameter in Eq. \ref{P approx}. The fit yields $\lambda_m = 15 \pm 2$ nm or $\epsilon_m = 3.7 \pm 0.7$, which reasonably compares with the estimate placing $\epsilon_m$ between 3.4 for dry polyamide~\cite{bason2007characterization} and 4.5, based on the Bruggeman approximation and 10\% swelling in water~\cite{bason2007characterization, freger2023dielectric}.  Mulhearn et al. concluded that the thinner films are likely to be less homogeneous and have a loose surface layer, which explains why they somewhat deviate from the present model and result in fairly large uncertainty of $\epsilon_m$. The increasing non-homogeneity of thinner films is also consistent with the moderate increase of the thickness-normalized water permeability, reported as well \cite{mulhearn2021thickness} (see Supplemental Material).

Intriguingly, similar to lipid layers, the physically reasonable value of $\epsilon_m$ yielded by fitting to the present model also rules out the alternative scenario that salt permeates polyamide as ion pairs  \cite{freger2020ion, freger2025pairing}, as it would produce a much weaker thickness dependence, similar to that of dipolar solutes such as water, urea, or glycerol \cite{paula1996permeation,paula1998permeation}. Instead, the observed thickness dependence indicates permeation primarily as single ions, which ensures maximal salt exclusion and may explain why aromatic polyamide films show exceptional selectivity in membrane desalination \cite{wijmans1995solution,elimelech2011future,geise2011water}.

We presume that this analysis may be extended to various synthetic films and membranes as well. Modern coating techniques, capable of producing highly uniform and conformal films of down to angstrom scale  with precise control and unprecedentedly low defect rates, such as atomic or molecular layer deposition \cite{gu_stafford2013molecular, yang_darling2018atomic} or layer-by-layer assembly \cite{bruening2008creation, regenspurg_deWos2024polyelectrolyte}, present an opportunity to follow this approach, as demonstrated here for both lipid and aromatic polyamide membranes. Remarkably, poor understanding of the physics behind $P_{\infty}$ is not an issue, as it is factored out of the thickness dependence, making the approach broadly applicable. Furthermore, deviations from the predicted thickness dependence, or unphysical fitted values of $\lambda_m$, may indicate alternative transport pathways, e.g., ion-pairs or more complex cooperative regimes, providing a tool for their differentiation from single-ion diffusion. This approach may thus benefit the design, optimization, characterization, and physical modeling of ultra-thin ion-selective films \cite{bannon2024application}.

\email{jay.singh@campus.technion.ac.il}
\email{mrk@technion.ac.il}
\email{vfreger@technion.ac.il}

\clearpage

\bibliographystyle{apsrev4-1}
\bibliography{refs}
\end{document}


\title{Supplemental Material \\ Ion Permeation in Nanoscale Films: Fundamental Limitation and Evaluation of Dielectric Properties}

\author{Jay Prakash Singh, Konstantin I. Morozov, and Viatcheslav Freger*}
\date{\today}
\maketitle

\section{Relation between energy profile across the membranes and its permeability}

Ion transport across the membrane is described within a steady-state continuum framework using the Nernst-Planck equation
\begin{equation}
\mathbf{J} = -D \left( \nabla C + C \nabla U \right),
\end{equation}
where $C(\mathbf{r})$ is the ion concentration, $D$ is the diffusivity inside the membrane, and $U(\mathbf{r})$ is the dimensionless electrostatic self-energy derived above. For transport normal to a planar membrane of thickness $d$, the problem reduces to one dimension,
\begin{equation}
J = -D \left( \frac{dC}{dz} + C \frac{dU}{dz} \right).
\end{equation}
Multiplying this equation by the integrating factor $e^{U(z)}$ and integrating across the membrane yields
\begin{equation}
J \frac{d}{D} \left\langle e^{U(z)} \right\rangle
= C_1 e^{U_1} - C_2 e^{U_2},
\end{equation}
where $\langle e^{U(z)} \rangle = d^{-1}\int_0^d e^{U(z)}\,dz$, and $C_{1,2}$ and $U_{1,2}$ denote concentrations and self-energies at the two interfaces.

The interfacial concentrations are related to the solution concentrations $(C_s)_{1,2}$ through
\begin{equation}
C_{1,2} = (C_s)_{1,2} \, e^{-(U_0 + U_{1,2})},
\end{equation}
where $U_0$ is a thickness-independent intrinsic contribution accounting for short-range effects such as ion solvation, steric exclusion, and molecular friction. Assuming symmetric boundary conditions ($U_1 = U_2$), the permeability is defined as
\begin{equation}
P(d) = \frac{J d}{C_1 - C_2}.
\end{equation}
Combining the above relations leads to the general result
\begin{equation}
P(d) = \frac{P_\infty}{\langle e^{U(z)} \rangle},
\end{equation}

where $P_\infty = D e^{-U_0}$ is the permeability of an infinitely thick membrane. All thickness dependence of the permeability therefore arises solely from the spatially averaged electrostatic self-energy profile $U(z)$ derived in the preceding sections. Substituting the approximate analytical form of $U(z)$ obtained above yields the closed-form expression for the thickness dependence of permeability reported in the main text. 

\section{Derivation of thickness-dependent part of the self-energy $U(z)$}
We consider a point charge $+e$ embedded in a planar dielectric slab (the membrane) of thickness $d$ and permittivity $\epsilon_m$, surrounded on both sides by water of permittivity $\epsilon_w$. Due to the dielectric mismatch, the charge induces polarization charges at the slab-water interfaces. To describe the resulting electrostatic potential $\phi$, it is convenient to work in cylindrical coordinates $(r,z)$ and expand the solution in Fourier-Bessel modes, which naturally incorporate the boundary conditions at the interfaces. 

We position the planar memrbane-water interfaces at $z=0$ and $z=d$ and place the charge at $r=0$ and $z=z_0\in(0,d)$. We will seek electric potential in the following general form: Within the membrane ($0<z<d$), it can be written as \cite{BatyginToptygin78}

\begin{figure}
    \centering
    \includegraphics[width=0.5\linewidth]{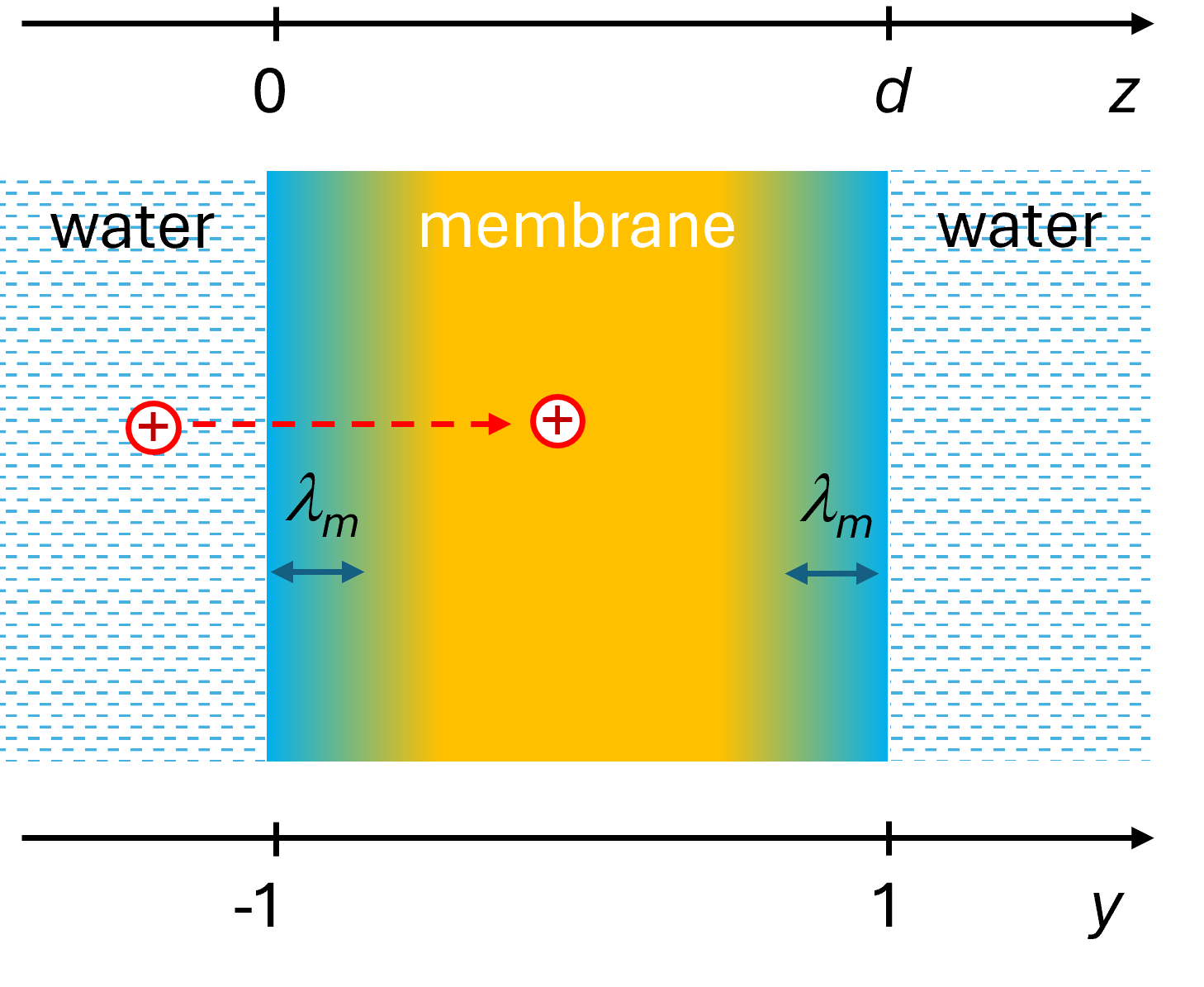}
    \caption{ Sketch of an ion crossing a planar film (membrane) of thickness $d$. $z$ (in units of length) and $y$ (dimensionless) are two alternative thickness coordinates used in the derivation. $\lambda_m$ represent the characteristic thickness of regions adjacent to the interfaces where ion exclusion weakens. }
    \label{fig:cordinate}
\end{figure}

\begin{equation}
	\phi_m(r,z) 
	= \frac{e}{4\pi\epsilon_0\epsilon_m} \int_0^\infty j_0(kr)\,e^{-k|z-z_0|}\,dk 
	+ \int_0^\infty j_0(kr)\!\left[ A(k)\,e^{k z} + B(k)\,e^{-k z} \right] dk,
	\label{eq:phi-m}
\end{equation}
where $j_0(x)$ is the zeroth-order Bessel function of the first kind, and $A(k), B(k)$ are coefficients determined by the boundary conditions at $z=0$ and $z=d$. The first term is the bare field produced by the point charge, and the unknown amplitudes $A(k)$ and $B(k)$ are produced by reflections from the two interfaces. In the water regions, only the decaying exponentials are allowed, thereby the potential may be written
\begin{align}
	\phi_w^-(r,z) &= \int_0^\infty j_0(kr)\,C_-(k)\,e^{k z}\,dk, \qquad z<0, \label{eq:phi-w-} \\
	\phi_w^+(r,z) &= \int_0^\infty j_0(kr)\,C_+(k)\,e^{-k (z-d)}\,dk, \qquad z>d,
	\label{eq:phi-w+}
\end{align}
where $C_\pm(k)$ are determined by the boundary conditions.

Since $j_0(kr)$ forms an orthogonal basis, the standard interface conditions apply \emph{for each $k$ separately}:
(i) Continuity of the potential, $\phi$, and (ii) continuity of the normal derivation of the displacement $\epsilon\,\partial_z \phi$ at $z=0$ and $z=d$. This yields four linear equations
\begin{equation}
\begin{aligned}
&\text{at } z=0: &&
\frac{e}{4\pi\epsilon_0\epsilon_m} e^{-k z_0} + A + B \;=\; C_-, \\
&&& \epsilon_m\!\left[ \frac{e}{4\pi\epsilon_0\epsilon_m} \,e^{-k z_0} +  A -  B \right]
\;=\; \epsilon_w C_-, \\[6pt]
&\text{at } z=d: &&
\frac{e}{4\pi\epsilon_0\epsilon_m} e^{-k (d-z_0)} + A e^{k d} + B e^{-k d}
\;=\; C_+, \\
&&& \epsilon_m\!\left[ \frac{-e}{4\pi\epsilon_0\epsilon_m} \,e^{-k (d-z_0)} + A e^{k d} - B e^{-k d} \right]
\;=\; -\,\epsilon_w  C_+.
\end{aligned}
\label{eq:BCs}
\end{equation}

The signs in the derivative terms follow from $\partial_z e^{\pm kz} = \pm k e^{\pm kz}$ and the outward normals at $z=0$ and $z=d$.

From \eqref{eq:BCs} we can eliminate  $C_-$ and $C_+$ by expressing them in terms of $A,B$. After collecting like terms, we obtain a $2\times 2$ linear system for $A$ and $B$:
\begin{equation}
\begin{aligned}
(\epsilon_m - \epsilon_w) A - (\epsilon_m + \epsilon_w) B
&=- \frac{e}{4\pi\epsilon_0\epsilon_m} (\epsilon_m - \epsilon_w)\, e^{-k z_0}, \\[6pt]
(\epsilon_m + \epsilon_w) A e^{k d} - (\epsilon_m - \epsilon_w) B e^{-k d}
&= \,\frac{e}{4\pi\epsilon_0\epsilon_m} (\epsilon_m - \epsilon_w)\, e^{-k (d-z_0)} .
\end{aligned}
\label{eq:linear-system}
\end{equation}

from which we ultimately obtain, 
\begin{align}
A(k) &=- \frac{e \Delta}{4 \pi \epsilon_0 \epsilon_m} \, e^{-k z_0} \;
\frac{ e^{-2 k(d- z_0)}-\Delta e^{-2 k d} }{ 1 - \Delta^2 e^{-2 k d} }, \\[1mm]
B(k) &=- \frac{e \Delta}{4 \pi \epsilon_0 \epsilon_m} \, e^{k z_0} \;
\frac{e^{-2 k z_0}-\Delta e^{-2 k d} }{ 1 - \Delta^2 e^{-2 k d} }.
\end{align}

where we introduced the dielectric contrast $\Delta = {({\epsilon_w - \epsilon_m})/({\epsilon_w + \epsilon_m})}$. In these equations, the denominators sum up multiple reflections, while the exponentials in the numerators encode the infinite sequence of image charges. 

These expressions may be substituted back to eq. \eqref{eq:phi-m} to calculate the potential $\phi_m(r,z)$, which may be used to compute the full electrostatic self-energy as $\tfrac{1}{2} e \phi_m(0,z_0)$. However, since we are interested in its thickness-dependent part only, we note that it is contained only in the second integral of eq. \eqref{eq:phi-m}. Therefore the required part of the self energy, here and below scaled to $k_BT$, is found as follow

\begin{equation}
	U(z_0) = \frac{e}{2k_BT}\int_0^\infty \!\!\left[A(k)\,e^{k z_0} + B(k)\,e^{-k z_0}\right] dk={-}\frac{\lambda_m \Delta}{2} \int_0^\infty 
	\frac{e^{-2k z_0} + e^{-2k(d-z_0)} {-} 2\Delta e^{-2kd}}
	{1 - \Delta^2 e^{-2kd}}\, dk,
	\label{U-final}
\end{equation}
where we {used} the membrane Bjerrum length
\begin{equation}
	\lambda_m \equiv \frac{e^2}{4\pi\epsilon_0\epsilon_m k_B T}.
\end{equation}

\section{Expanded forms of $U(z)$}

Expanding the denominator of eq. \eqref{U-final} as a geometric series, we obtain
\[
\frac{1}{1 - \Delta^2 e^{-2kd}} = \sum_{n=0}^\infty (\Delta^2)^n e^{-2knd},
\]
and, after evaluating the $k$-integrals and dropping 0 subscript in $z_0$, we have 

\begin{equation} \label{U_series}
	U(z) = {-}\frac{\lambda_m}{4} \sum_{n=0}^{\infty} 
	\left[ 
	\frac{\Delta^{2n+1}}{z + nd} + 
	\frac{\Delta^{2n+1}}{d-z + nd} {-}
	\frac{2}{d} \frac{(\Delta^2)^{n+1}}{n+1} 
	\right].
\end{equation}
The first two terms represent interactions with the lower and upper interfaces (and their multiple reflections), while the third is a $z$-independent background.

Towards approximation, it is convenient to replace first $z$ with a symmetric variable as shown in Fig. \ref{fig:cordinate}

\[
y = \frac{2z}{d} - 1, \qquad y \in [-1,1],
\]
so that $z=\tfrac{d}{2}(1+y)$. Expanding the first two terms in brackets of eq. \eqref{U_series} around $y=0$, whereby only even powers survive, yields

\begin{equation}
   U(y)={-}\frac{\lambda_m}{2d}\sum_{n=0}^{\infty}\left[\frac{\Delta^{2n+1}}{2n+1+y} + \frac{\Delta^{2n+1}}{2n+1-y} {-} \frac{{{(\Delta^2)}}^{n+1}}{n+1}\right] = {-}\frac{\lambda_m}{2d} \sum_{n=0}^{\infty}\left[2\frac{\Delta^{2n+1}}{2n+1}+ 2\sum_{l=1}^{\infty}\frac{\Delta^{2n+1}}{(2n+1)^{2l+1}}y^{2l}{-}\frac{(\Delta^2)^{n+1}}{n+1} \right].
\end{equation}

Using for the first term the relation   
\[
2\sum_{n=0}^{\infty}\frac{x^{2n+1}}{2n+1}=\ln\frac{1+x}{1-x},\]
for the last term the relation 
\[\sum_{n=0}^\infty \frac{x^{n+1}}{n+1} = -\ln(1-x),\]
and changing the order of summation for the second term, we obtain an alternative exact $y$-expansion for the thickness-dependent part of the self-energy

\begin{equation} \label{U_y}
    U(y)={-}\frac{\lambda_m}{2d}\left[\ln\frac{1+\Delta}{1-\Delta}+2\sum_{l=1}^{\infty}\sum_{n=0}^{\infty}\frac{\Delta^{2n+1}}{(2n+1)^{2l+1}}y^{2l}{+}\ln(1-\Delta^2)\right]= {-}\frac{\lambda_m}{d} \left[ \ln(1{+}\Delta) + \sum_{l=1}^{\infty} y^{2l}\,\chi_{2l+1}(\Delta) \right]
\end{equation}

where coefficients
\begin{equation} \label{chi_def}
	\chi_{m}(\Delta) \equiv \sum_{n=0}^{\infty} \frac{\Delta^{2n+1}}{(2n+1)^m},
\end{equation}
are known as the Legendre chi functions of order $m$.

\section{Approximations for $U(z)$ and thickness dependence of permeability}
For  {$0<\Delta<1$}, the coefficients $\chi_{2l+1}(\Delta)$ are well approximated by $\Delta$, i.e.
\[
\chi_{2l+1}(\Delta) \approx \Delta, \quad l \geq 1.
\]
The last sum in eq. \eqref{U_y} is then approximated as the sum of a geometric series. Ultimately, eq. \eqref{U_y}  is reduced to the compact approximate form

\begin{equation} \label{U_approx}
	U(y) \approx {-}\frac{\lambda_m}{d} \left[ \ln(1{+}\Delta) + \Delta \frac{y^2}{1-y^2} \right].
\end{equation}

The thickness-dependence of permeability is then obtained as

\begin{equation} \label{P_approx}
	\frac{P(d)}{P_{\infty}} 
    = \left(\tfrac12 \int_{-1}^1 e^{U(y)}dy\right)^{-1}
   \approx 2\,(1 {+} \Delta)^{\tfrac{\lambda_m}{d}}
	\left( \int_{-1}^{1} \exp\left[ {-}\Delta\frac{\lambda_m}{d}\frac{y^2}{1-y^2} \right] dy \right)^{-1} = 
(1 {+} \Delta )^{\tfrac{\lambda_m}{d} } {\beta\left(\frac {\Delta}{2}\frac{\lambda_m}{d} \right)},
\end{equation}
where $\beta(x)=\{ x{e^{x}}[K_1(x)-K_0(x)]\}^{-1}$, and $K_0$ and $K_1$ are the modified Bessel functions of the zero and first order, respectively.

\begin{figure}
    \centering
    \includegraphics[width=0.5\linewidth]{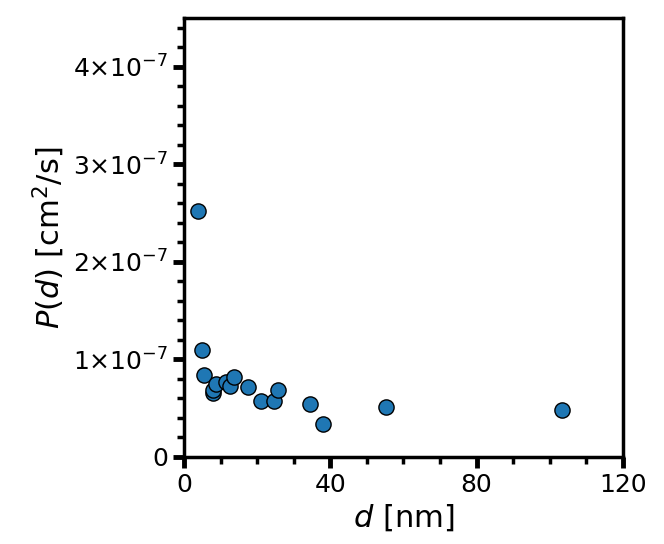}
    \caption{ Thickness-normalized water permeability versus nominal membrane thickness of analyzed polyamide films, adopted from Ref.~\cite{mulhearn2021thickness}. }
    \label{fig:water_correction}
\end{figure}

\section{Water permeability of analyzed polyamide film (Mulhearn et al)}

For comparison with NaCl permeability reported in Ref.~\cite{mulhearn2021thickness}, we assume that all films are homogeneous and have an identical chemical structure, in which case the water permeability would remain constant. However, as shown in Figure S2, at lower membrane thicknesses the water permeability deviates from the nominal constant value, highlighting the likely non-homogeneity of membrane and possible presence of a looser surface layer that more strongly affects the thinner membranes.

\bibliography{refs}  